 \documentclass{emulateapj}

\newcommand{\numax}{\mbox{$\nu_{\rm max}$}}
\newcommand{\Dnu}{\mbox{$\Delta \nu$}}

\newcommand{\muHz}{\mbox{$\mu$Hz}}
\newcommand{\most}{\textit{MOST}}

\slugcomment{accepted for publication in ApJ}

\shorttitle{Oscillations and activity in Procyon}
\shortauthors{D. Huber et al.}

\begin{document}

\title{Solar-like oscillations and activity in Procyon: A comparison of the 2007 
\textit{MOST}\altaffilmark{*} and
ground-based radial velocity campaigns}

\author{
Daniel Huber\altaffilmark{1},
Timothy R. Bedding\altaffilmark{1},
Torben Arentoft\altaffilmark{2},
Michael Gruberbauer\altaffilmark{3},
David B. Guenther\altaffilmark{3},
G\"unter Houdek\altaffilmark{4},
Thomas Kallinger\altaffilmark{4,5},
Hans Kjeldsen\altaffilmark{2},
Jaymie M. Matthews\altaffilmark{5},
Dennis Stello\altaffilmark{1}, and 
Werner W. Weiss\altaffilmark{4}}

\altaffiltext{*}{Based on data from the MOST satellite, a Canadian Space Agency mission,
jointly operated by Dynacon Inc., the University of Toronto Institute for Aerospace Studies and the 
University of British Columbia, with the assistance of the University of Vienna.}
\altaffiltext{1}{Sydney Institute for Astronomy (SIfA), School of Physics, University of Sydney, NSW 2006, Australia; \mbox{dhuber@physics.usyd.edu.au}}
\altaffiltext{2}{Danish AsteroSeismology Centre (DASC), Department of Physics and Astronomy, Aarhus 
University, DK-8000 Aarhus C, Denmark}
\altaffiltext{3}{Institute for Computational Astrophysics, Department of Astronomy and Physics, 
Saint Marys University, NS B3H 3C3, Halifax, Canada}
\altaffiltext{4}{University of Vienna, Institute for Astronomy, 1180 Vienna, Austria}
\altaffiltext{5}{Department of Physics and Astronomy, University of British Columbia, 6224 Agricultural Road 
Vancouver, BC V6T 1Z1, Canada}

\begin{abstract}
We compare the simultaneous 2007 space-based \most\ photometry and ground-based 
radial velocity observations of the F5 star Procyon. We identify slow variations in the \most\ data 
that are similar to those reported in the radial velocity (RV) time series, and confirm 
by comparison with the Sun that these variations are likely the signature of stellar activity. 
The \most\ power spectrum yields clear evidence for individual oscillation frequencies that match 
those found in the radial velocity data by \citet{bedding10}. We identify the same ridges due to modes of different 
spherical degree in both datasets, but are not able to confirm a definite ridge identification 
using the \most\ data. We measure the luminosity amplitude per radial mode 
$A_{l=0,\rm phot}= 9.1\pm0.5$\,ppm. Combined with the estimate for the RV data by \citet{arentoft} this 
gives a mean amplitude ratio of $A_{l=0,\rm phot}/A_{l=0,\rm RV} = 0.24 \pm 0.02$\,ppm\,
cm$^{-1}$\,s, 
considerably higher than expected from scaling relations but in reasonable agreement with theoretical models by 
\citet{houdek09}. We also compare the amplitude ratio as a function of frequency, and find 
that the maximum of the oscillation envelope is shifted to higher frequencies in photometry than 
in velocity.
\end{abstract}


\section{Introduction}
The detection and measurement of oscillations in stars provides a unique possibility to infer 
details about the physics governing their interiors. The prospect of extending such studies 
from the Sun to distant stars has motivated many observation campaigns in recent 
decades. Even with the wealth of new space-based photometry from \textit{CoRoT} 
\citep[see, e.g.,][]{michel08} and \textit{Kepler} \citep[see, e.g.,][]{gilliland} 
there is still an important place for ground-based spectroscopic campaigns of bright nearby 
stars with well-known fundamental parameters. Owing to its brightness ($V=0.3$), proximity (d = 3.5\,pc), 
and membership in an 
astrometrically well determined binary system, the F5 sub-giant Procyon A ($\alpha$\,CMi, HR\,2943, 
HD\,61421) has long been considered a prime target for such campaigns. 

The majority of early efforts to detect oscillations in Procyon have relied on measuring Doppler 
velocities from a single site. The first claimed detection dates back to \citet{gelly} which, however, 
could not be 
confirmed by \citet{libbrecht} or \citet{innis}, who reported null-detections at similar 
sensitivity levels. With the benefit of hindsight, it now seems that the first detection of 
power excess in Procyon (and, in fact, any 
other solar-like star than the Sun) was by \citet{brown}. This was followed by numerous observing campaigns, 
mostly single-site, 
taking advantage of the increasing precision of Doppler-shift measurements \citep{mosser,
barban99,martic99,martic04,claudi,eggenberger,bouchy04,leccia}. While all of these studies revealed clear power excess 
in the expected frequency range of 0.5--1\,mHz, a consistent determination of individual frequencies 
was hampered due to severe aliasing caused by daily gaps inevitable in dual or single-site 
observations.

The first two sets of continuous observations of Procyon by the Canadian \most\ satellite 
\citep{walker,matthews} in 2004 and 2005 resulted in null-detections, leading to the conclusion 
that luminosity amplitudes in Procyon must be lower than 15\,ppm and/or the mode lifetimes 
shorter than 2--3 days \citep{matthews04,guenther07}. 
\citet{bedding05} found these results to be compatible with 
limits set from ground-based radial velocity observations and \citet{baudin} confirmed the null-result, 
while \citet{regulo} and \citet{marchenko} cautiously claimed a detection of oscillations based on a re-analysis of \most\ 
data. Meanwhile, \citet{bruntt} reported a detection of power excess with amplitudes of about 
twice the solar value ($\sim$ 8\,ppm) based on continuous space-based photometry by the 
\textit{WIRE} satellite, consistent with the upper limits set by the \most\ results.

The clear identification of individual oscillation modes in Procyon was finally achieved with a large 
ground-based radial velocity campaign that was carried out in January 2007 \citep{arentoft,bedding10}. 
Simultaneously, a third set of \most\ observations, longer and with higher precision than the 
previous runs, was obtained in January and February 2007 \citep{guenther08}.
Here we present the first direct comparison of these datasets.

\section{Summary of observing campaigns}

The following sections present a brief summary of the observations and main results of the 
two campaigns on which the comparison in this paper is based. For a brief introduction to 
basic characteristics of solar-like oscillations, relevant to Procyon, we refer the reader 
to Section 2 in \citet{bedding10}.

\subsection{\most\ photometry}
The \most\ (Microvariability and Oscillations of STars) space telescope, launched in 2003, is the 
first satellite dedicated to asteroseismic observations from space. \most\ houses a 
15\,cm telescope, with observations performed through a custom broad-band filter (350--700\,nm). It 
is positioned in a sun-synchronous low-Earth orbit, enabling it to continuously monitor stars for up 
to three months. For a detailed description of the instrument, we refer to \citet{walker}.

\citet{guenther08} presented the results of the \most\ 2007 campaign. The data are the most precise 
\most\ Procyon photometry to date, outperforming the previous runs in both length (38.5\,d) and 
time-series point-to-point scatter (140\,ppm), and consequently also in high-frequency noise (0.9\,ppm). 
While the power spectrum showed an excess in the expected oscillation frequency range and 
an autocorrelation of the spectrum yielded strong evidence for the expected characteristic large 
frequency separation (\Dnu) of $\sim$\,55\muHz, no reliable individual mode frequencies could be 
extracted. Using various common and new frequency extraction tools, \citet{guenther08} found that 
no consistent regularly spaced frequencies could be identified unless it was assumed that the 
mode lifetimes are $\sim$\,2 days, i.e. slightly less than solar \citep[see, e.g.,][]{chaplin_tau}. 
The large scatter of the extracted frequencies around the predicted 
regularity ($\pm$5\,\muHz) was found to be consistent with the scatter of the frequency 
identifications from previous radial velocity campaigns. Using new numerical convection models 
\citet{guenther08} argued that, contrary to the Sun, the granulation timescale in Procyon 
is similar to the timescale of the p-mode oscillations. This was identified as a possible reason 
for the short mode lifetimes in Procyon and the consequent difficulty of extracting consistent 
frequencies from the \most\ 2007 data.

\subsection{Ground-based radial velocity campaign}
The 2007 ground-based radial velocity (RV) campaign was described in detail by \citet{arentoft} and included 
11 telescopes, with apertures ranging from 0.6 to 3.9\,m, at eight observatories. Covering a total 
length of 25\,days, with a duty cycle above 90\% for the central 10 days of observations, it is the 
most precise and complete radial-velocity campaign dedicated to asteroseismology to date.

\citet{arentoft} reported the detection of slow 
variations in the velocity timeseries with an apparent period of $\sim$10\,d, which they attributed 
to rotational modulation of active regions on the 
surface of the star. A possible rotation period of Procyon of $\sim$10\,d or twice that value 
was suggested, with the latter scenario being more likely if it is assumed that the rotation axis 
is aligned with the known 
inclination of the binary orbit. \citet{arentoft} also provided an estimate of the mean mode 
amplitude between $650-1150\muHz$ of 38.1$\pm$1.3 cm\,s$^{-1}$, consistent with previously reported 
detections and upper limits.

A detailed asteroseismic analysis of the RV data was presented by \citet{bedding10}. The 
continuous coverage and high S/N allowed the first measurements of mode 
frequencies in Procyon. These authors also measured large and small frequency separations as a function of 
frequency and identified a possible mixed mode at 446\,\muHz. They used the variation of peak 
amplitudes caused by the stochastic nature of the oscillations to estimate the mode lifetimes in Procyon 
to be 1.3$\pm$0.5\,d. As a result of the large linewidths of the modes, however, a major difficulty in 
the analysis was the identification of modes with even and odd spherical degree $l$. \citet{bedding10} 
gave several arguments for the most probable mode identification but left open the possibility that 
the identification could be reversed.

\section{Slow variations}

\begin{figure}
\begin{center}
\resizebox{\hsize}{!}{\includegraphics{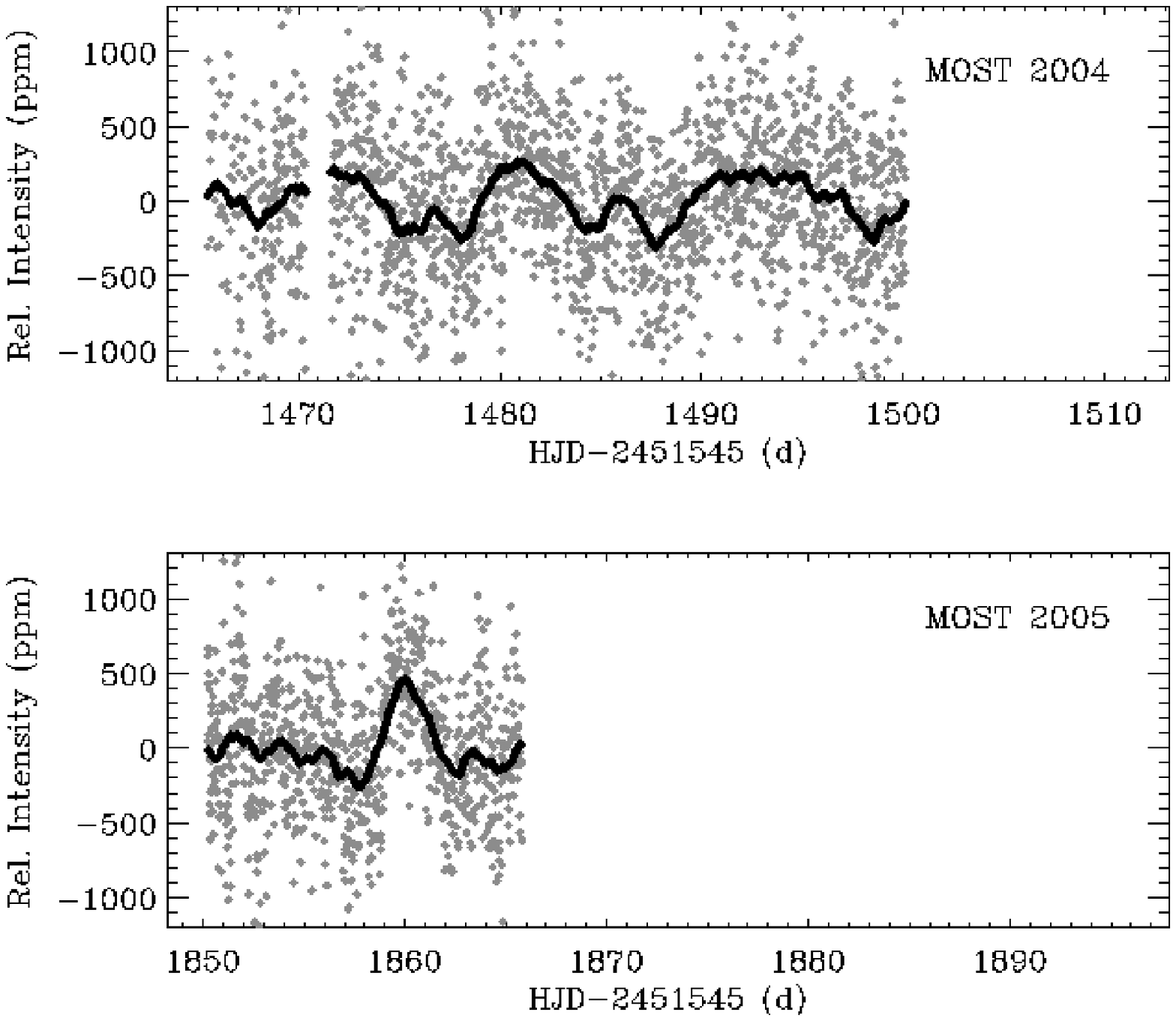}}
\resizebox{\hsize}{!}{\includegraphics{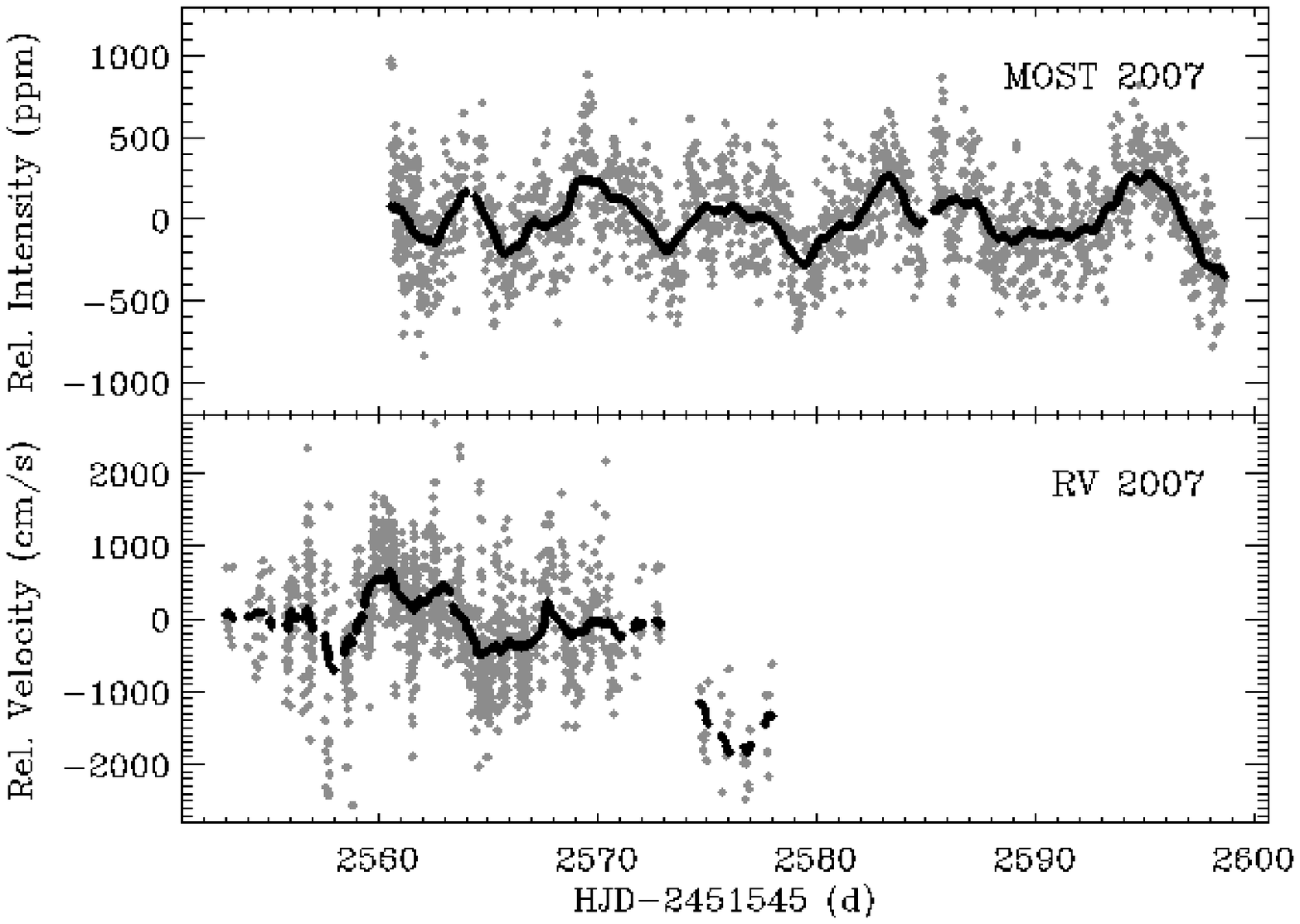}}
\caption{Detrended \most\ light curves of Procyon obtained in 2004, 2005 and 2007 (every 50th datapoint shown) 
and 2007 radial velocity curve (every 20th datapoint shown) of Procyon. The thick black lines show the 
result of smoothing with a boxcar of width $\sim$ 2 days.}
\label{fig01}
\end{center}
\end{figure}

The Procyon \most\ photometry provides the opportunity to further investigate the nature of the slow 
variations detected in radial velocity. The stability and 
continuity of \most\ photometry has previously been successfully used to study activity in several 
stars \citep{croll06b,walker07}.

The analysis by \citet{guenther08} was based on a high-pass filtered \most\ light curve to 
focus on the detection of solar-like oscillations. Here, we use the raw light curve, as produced by the 
reduction pipeline described in \citet{reegen}, as the starting point of our analysis. 
As discussed by \citet{huber09a} for another solar-like star observed 
by \most\ (85\,Peg) the reduced photometry sometimes shows long-periodic instrumental variability that 
can be identified and corrected by decorrelating satellite telemetry data, such as board and 
preamplifier temperatures, against the observed target intensities. We applied the same technique to 
the 2004, 2005 and 2007 photometry of Procyon and the resulting detrended light curves are shown 
in Figure \ref{fig01}.

The intensity and velocity curves in Figure \ref{fig01} show 
similar variability in all datasets. The most prominent periodicity in the 
velocities, with a period of $\sim$10 days, is not readily apparent in the 2007 \most\ photometry, which 
shows the strongest signal as measured from the amplitude spectrum with a period of $\sim$6 days. 
However, as is well known for the Sun and demonstrated by \citet{clarke03} for models of 
spotted stars, the relationship between simultaneous velocity and intensity observations of active regions 
is not simple and strongly depends on parameters such as rotational velocity, 
inclination and spot size. Detailed modeling of the variations using the overlapping 
parts of the dataset ($\sim$ 10 days) is beyond the scope of this paper, but we make some 
qualitative statements on the variability based on the observed low frequency power levels.

\begin{figure}
\begin{center}
\resizebox{\hsize}{!}{\includegraphics{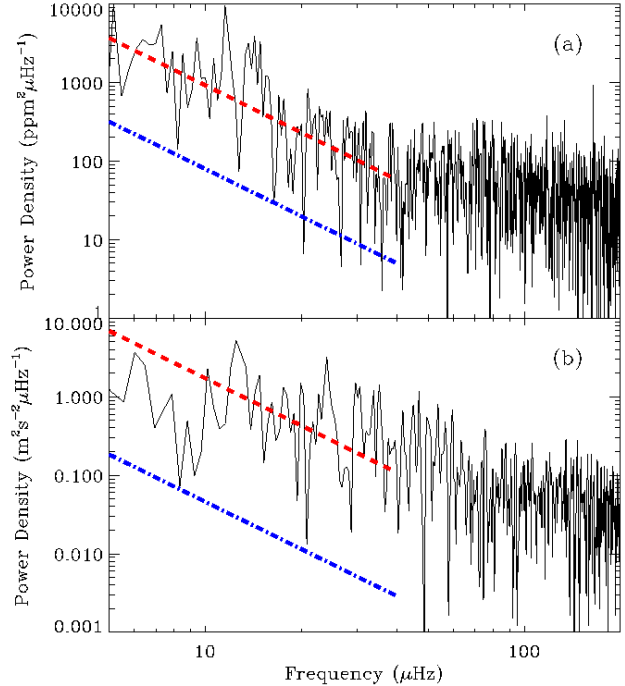}}
\caption{\textit{(a)} Photometric power density spectrum of Procyon. The red dashed line shows a 
power law with a fixed slope of 2 fitted in the frequency interval 5--40\,\muHz. The blue 
dashed-dotted line shows the average power density level of the Sun during solar maximum, 
determined using the same method with 30\,d subsets of SOHO data. \textit{(b)} Same as panel (a) but 
for radial velocities.}
\label{fig02}
\end{center}
\end{figure}

To compare both datasets independent of length and sampling, we converted the power spectra 
to power density by multiplying with the effective length of each dataset (calculated as the inverse 
of the area under the spectral window in power).
Figure \ref{fig02} shows the low-frequency power density level measured in Procyon in 
intensity (upper panel) and velocity (lower panel) using the full length of both datasets obtained 
in 2007. To measure the low frequency power density level, we fitted power laws with a fixed slope 
of 2 in the frequency interval 5--40\,\muHz\ (red dashed lines). For comparison, the blue dashed-dotted 
lines show the average power density level of the Sun during solar maximum derived using the same method 
with 30\,d subsets of data 
obtained in intensity by VIRGO \citep{frohlich} and in velocity by GOLF \citep{ulrich,garcia05}, 
both of which are instruments onboard the \textit{SOHO} spacecraft.
Note that we have corrected the 
photometric power density for the spectral response of the \most\ filter following 
the method of \citet{michel09}, yielding $R_{\rm g}=4.23$ using an effective temperature of 
$T_{\rm eff}$ = 6500\,K for Procyon, compared to 
$R_{\rm g}=5.02$ for solar VIRGO observations in the green channel. 

\begin{figure}
\begin{center}
\resizebox{\hsize}{!}{\includegraphics{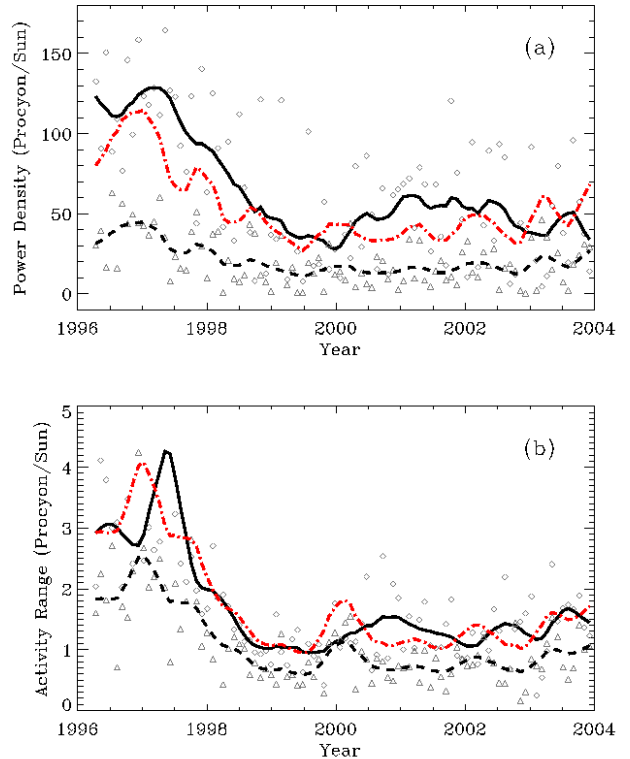}}
\caption{\textit{(a)} Ratio of low-frequency power density for Procyon and the Sun as a function 
of solar activity cycle in photometry (triangles, dashed line) and velocity 
(diamonds, solid line). Open symbols show individual measurements using 30-day subsets and 
lines are values which were smoothed twice using a boxcar with a width of $\sim$5 months. The red dashed-dotted line shows the 
photometric ratio scaled to account for the different 
projected rotational velocities of Procyon and the Sun. \textit{(b)} Same as Panel (a) but using the 
activity range as defined in \citet{basri}.}
\label{fig03}
\end{center}
\end{figure}

As noted by \citet{arentoft}, the Procyon velocity power spectrum shows a similar 1/$\nu^2$ dependence 
as observed for the Sun but at higher power density levels, and the same is observed in the 
photometry. Is the observed excess of photometric power density, measured relative to the Sun, 
consistent with the velocity observations? To investigate this, we used SOHO data spanning from 
1996 to 2004 and measured the low frequency power density levels in 
independent 30\,d subsets as described above
throughout the solar activity cycle. The ratios of the observed levels between Procyon and the Sun 
for each subset are 
shown in Figure \ref{fig03}(a).

We observe that the power density ratio between Procyon and the Sun is substantially higher in velocity than in 
photometry.  
\citet{arentoft} argued that the velocity power density at frequency $\nu$ is expected to scale as 
follows:

\begin{equation}
{\rm PD(\nu)_{RV}} = \left (\frac{da}{a} \right )^{2} \left (\frac{v \sin i}{T} \right )^{2} \nu^{-2} \: ,
\end{equation}

\noindent
where $da/a$ is the fractional area 
covered by active regions, $v \sin i$ the projected rotational velocity, and $T$ the typical 
lifetime of active regions on the stellar surface. 

Similarly, the amplitude measured in photometry will be proportional to $da/a$ 
and the luminosity variation $\delta L/L$ caused by the flux contrast between the unspotted 
and spotted areas of the star \citep[see, e.g.,][]{dorren}.
We can therefore rewrite Equation (1) for the case photometric power densities as:

\begin{equation}
{\rm PD(\nu)_{Phot}} = \left (\frac{da}{a} \right )^{2} \left (\frac{\delta L/L}{T} \right )^{2} \nu^{-2} \: .
\end{equation}

\noindent
We note that Equation (2) is only intended to give an approximate estimation of the 
photometric power density due to stellar activity. Compared to detailed photometric spot models 
\citep[see, e.g.][]{dorren,lanza03,mosser09}, Equation (2) for example does not include an explicit dependence on  
stellar inclination. As shown by \citet{mosser09}, this corresponds to neglecting any information 
about the latitude $\lambda$ of active regions (i.e., we assume in Equation (2) that it is equally 
likely to observe spots near the equator or near the poles). While this is certainly not the 
case for the Sun, this information is also neglected in Equation (1) (e.g., an active region at $\lambda=90^{\circ}$ on a 
star with $i=90^{\circ}$ will cause no velocity variation). Since we are here 
only interested in comparing the ratio of velocity to photometry variations, any explicit dependence 
on the latitude of active regions will therefore cancel out.

Equations (1) and (2) imply that the difference between the velocity 
and photometry power density ratio of two stars depends only on $v \sin i$ and $\delta L/L$. We assume that 
the ratio in $\delta L/L$ (and hence the flux contrast) between Procyon and the Sun is negligible 
compared to the ratio in $v \sin i$, which is supported by detailed 
spot modeling of stars hotter than the Sun \citep{lanza09,lanza11}. Using 
$v \sin i = 2.0$\,km\,s$^{-1}$ for the Sun and 
$v \sin i = 3.2$\,km\,s$^{-1}$ for Procyon \citep{allende}, this therefore implies a difference in the velocity 
and photometry power density ratios by a factor of $\sim 2.56$. The red dashed-dotted line in Figure 
\ref{fig03}(a) shows the photometric power density ratio multiplied by this value. We observe that 
the power density ratios are now in better agreement, both during solar minimum (in $\sim$1996) and 
solar maximum (in $\sim$2002). This 
result implies that the observed variations of Procyon in both datasets are qualitatively in 
agreement with being the signature of stellar activity.

The scatter of individual datapoints in Figure \ref{fig03}(a) is large, presumably 
due to the complex non-sinusoidal variations of stellar activity causing large variations in 
low frequency power density fits. We have therefore repeated the above exercise using the activity range as 
defined in \citet{basri}, which measures the maximum absolute deviation of the time series with 
respect to its mean. The 
results of this are shown in Figure \ref{fig03}(b). In this case, the scaling factor is expected
to be $\sqrt{2.56}$. Again, the scaled photometric ratios are in reasonable agreement with the 
velocity ratios, confirming the results derived using power densities.

\vspace{0.5cm}
\section{Solar-like oscillations}

\subsection{Comparison of power spectra}

\begin{figure*}
\resizebox{\hsize}{!}{\includegraphics{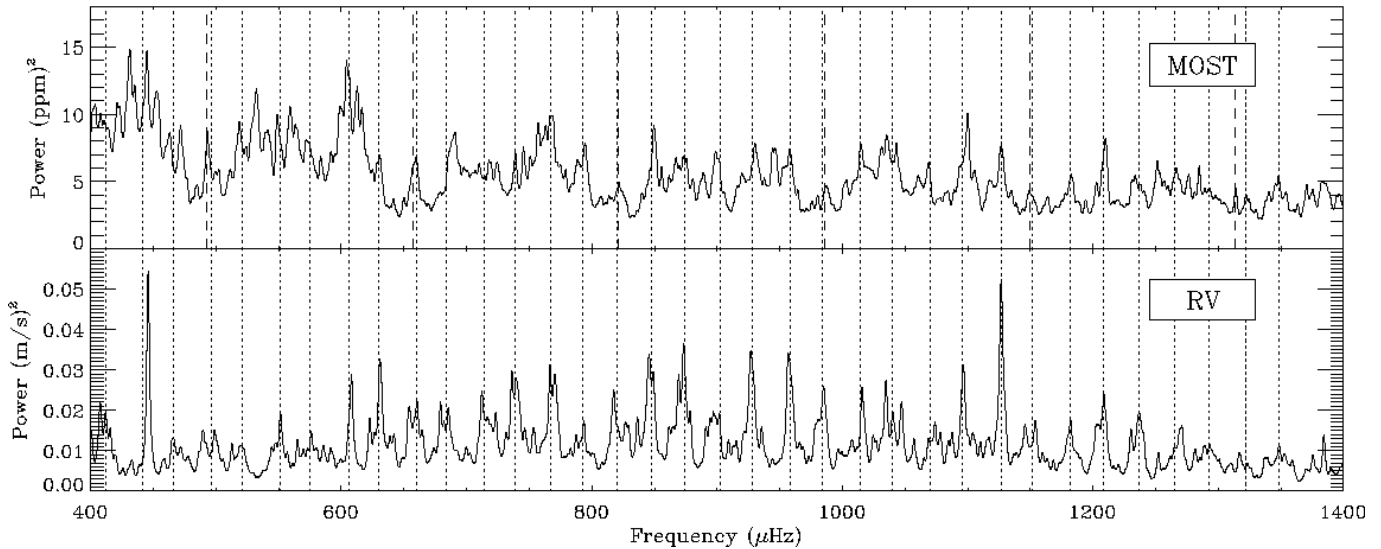}}
\caption{Power spectra of Procyon from \most\ photometry (upper panel) and 
the radial velocity campaign (lower panel), smoothed with a Lorentzian with a width of 2.5\muHz\ 
(corresponding to a mode 
lifetime of 1.5\,d). Dotted lines mark the odd and even ridge centroids identified in the 
velocity data. Dashed lines show the harmonics of the \most\ orbital frequency.}
\label{fig04}
\end{figure*}

Figure \ref{fig04} compares the power spectra of both full 2007 datasets 
in the frequency range where p modes have been detected. We used the sidelobe-optimized 
power spectrum for the velocities, as described by \citet{bedding10}, and the high-pass filtered \most\ 
data used by \citet{guenther08}. Note that the funnels of low power around the orbital harmonics of \most\ 
(marked as dashed lines) are due to the high-pass filter removing power leaking from low frequencies, 
which in turn is caused by periodic outlier rejections in the data reduction pipeline during high-straylight 
phases. To smooth over the effects caused by the stochastic nature of the oscillations, we 
convolved both spectra with a Lorentzian profile with a width of 2.5\muHz,
corresponding to a mode lifetime of 1.5\,d. We also show with dotted lines the values for the odd and 
even ridge centroids determined by \citet{bedding10} for the velocity data (see their Figure 9). 
Note that although our adopted value for the mode lifetime is larger than measured for 
other F-stars such as HD\,49933 \citep{gruberbauer09,benomar}, the exact choice of this value has 
no influence on the results presented below.

The comparison shows clearly that most of the peaks in the \most\ spectrum coincide with 
pulsation frequencies identified in the velocity power spectrum. The agreement 
improves considerably towards high frequencies ($>800$\muHz), where the 
granulation background in photometry becomes lower, as can be seen by the steady 
decrease of power towards high frequencies (see also Section \ref{sec:amps}). We also see 
that, unluckily, several of the intrinsic pulsation frequencies of Procyon near 
maximum power coincide almost exactly with harmonics of the orbital frequency of the \most\ satellite. 

\begin{figure}
\begin{center}
\resizebox{\hsize}{!}{\includegraphics{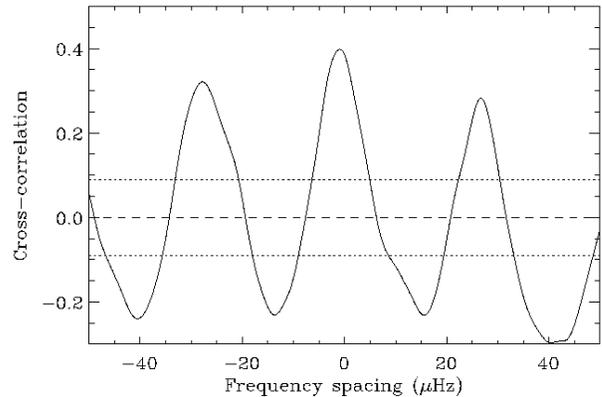}}
\caption{Cross-correlation of the velocity and photometry power spectra in the frequency range 
650-1150$\,\mu$Hz. The dashed and dotted lines mark the mean and $\pm 1\sigma$ noise level derived 
from cross-correlating the 
RV power spectrum with 2000 \most\ power spectra computed from white noise time 
series with the same sampling and scatter as the original data.}
\label{fig05}
\end{center}
\end{figure}

To investigate the agreement between the two spectra 
in more detail, we calculated a cross-correlation of the two power spectra in the 
central region of maximum power from 650-1150$\,\mu$Hz. The result is shown in Figure 
\ref{fig05}. As expected, we see clear 
maxima at zero offset and at half the large frequency separation. Note that the shift of the highest 
peak from zero offset is only 0.9\,\muHz, which is much smaller than the lifetime of the modes and 
therefore insignificant. To test the significance of the peak height, we correlated the RV power 
spectrum with \most\ power spectra calculated from 
2000 white noise time series with the same sampling and scatter as the original dataset. The 
resulting distribution at zero offset showed a mean of zero and a standard deviation of 0.09, which is 
indicated as dashed and dotted lines in Figure \ref{fig05}. At a level of $>4\sigma$, 
these results confirm that the peaks observed in the \most\ power spectrum very likely correspond to the 
oscillations observed in the radial velocity data.

\subsection{\'{E}chelle diagrams and folded power spectra}

\begin{figure*}
\begin{center}
\resizebox{16cm}{!}{\includegraphics{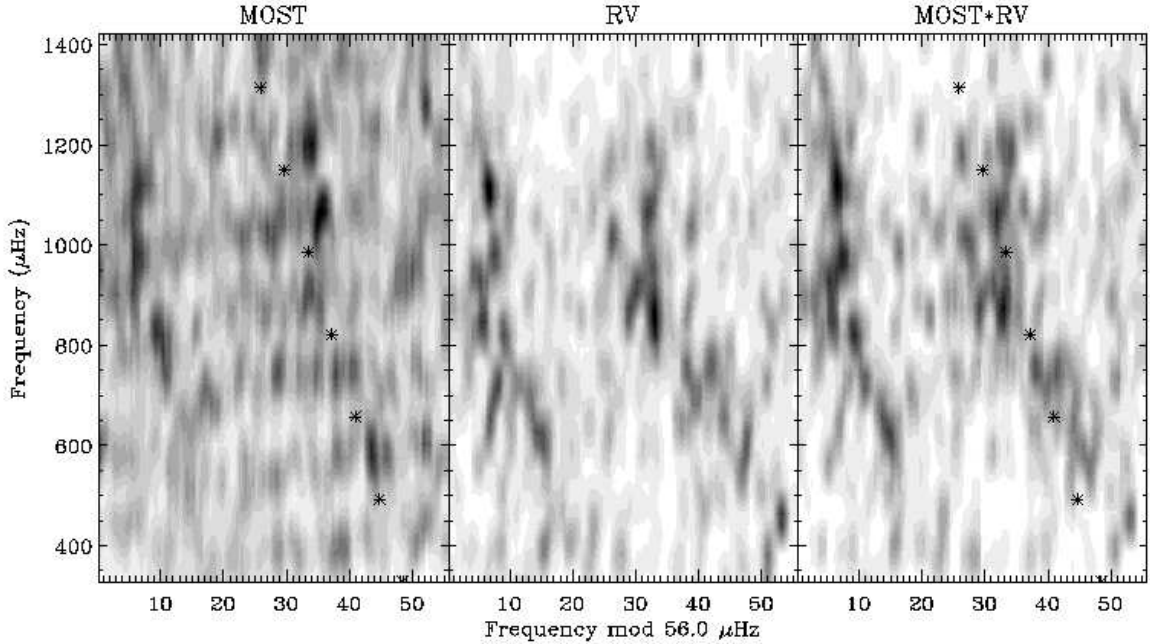}}
\caption{\'{E}chelle diagrams of \most\ (left panel) and radial velocity (middle panel) power 
spectra, as well as the combination of both data sets (right panel). Darker 
greytones correspond to higher power. The asterisks in the left and right panel mark the harmonics 
of the \most\ orbital frequency.}
\label{fig06}
\end{center}
\end{figure*}

A widely used method to analyze the regular frequency pattern characterizing solar-like oscillations 
is to stack the power spectrum (or extracted frequencies) in slices of the large frequency 
separation \Dnu, forming a so-called \'{e}chelle diagram \citep{grec}. 
Note that throughout the paper we use $\Dnu=56\,\muHz$, which corresponds to the large 
separation at maximum power as identified in the RV data \citep[see Figure 11(a) in][]{bedding10}.
The left and middle panel of Figure \ref{fig06} show \'{e}chelle diagrams 
of both power spectra smoothed to the same frequency resolution.
Note that the \most\ power spectrum has been corrected for the 
background contribution due to granulation and activity (see Section \ref{sec:amps}). The \'{e}chelle 
diagrams clearly show two ridges corresponding to modes of odd and even degree in both datasets. The 
similarity of the curvature of both ridges (and hence the variation of \Dnu\ with frequency) in 
both individual datasets reaffirms our conclusion that the peaks seen in Figure \ref{fig04} are intrinsic 
p modes. The \'{e}chelle diagrams also show that the higher low-frequency noise in the photometry 
makes it harder to detect p modes below $\sim800$\muHz\ than in the velocity data.

In addition to the \most\ and RV datasets, we also analyzed a power spectrum constructed by 
multipliying both individual power spectra. This corresponds to a power 
spectrum of the convolution of the \most\ and RV timeseries, combining 
the advantage of the higher frequency resolution of the \most\ data with the higher S/N of the RV 
data. The resulting \'{e}chelle diagram displayed in the right panel of Figure \ref{fig06} clearly 
shows more well-defined ridges than the corresponding diagrams using the individual datasets.

Having confirmed that we have detected the ridges in both datasets, an obvious step in the analysis 
is to attempt to confirm or discard the ridge identification presented in \citet{bedding10} using the 
\most\ data.
It is well known that observations in intensity are less sensitive to pulsation modes of higher 
spherical degree than velocity measurements. We therefore 
expect ridges observed in photometry to be shifted to higher frequencies in the 
\'{e}chelle diagram than in velocity, with the amount of shifting depending on the ridge 
identification. Calculations based on the theoretical response functions by \citet{kjeldsen08} and 
the frequency values presented in \citet{bedding10}, however, showed that the expected shift is only of the 
order of 1\,\muHz, too small to be detected with the current uncertainties.

\begin{figure}
\begin{center}
\resizebox{\hsize}{!}{\includegraphics{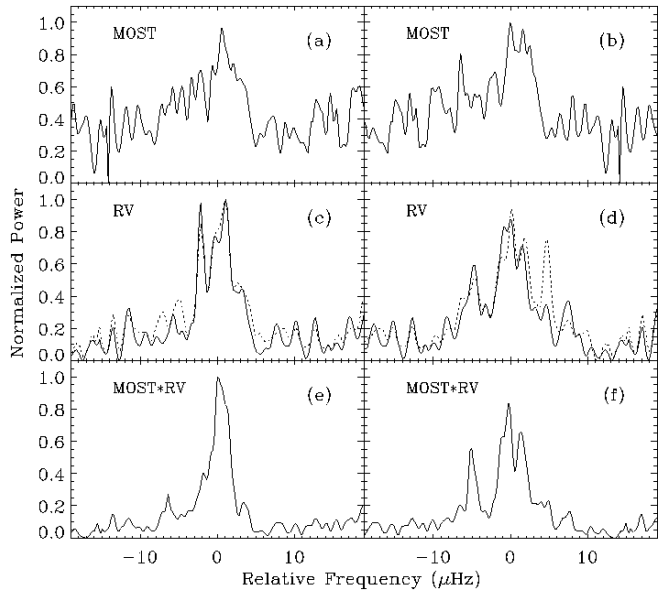}}
\caption{Collapsed \'{e}chelle diagrams in the region $700-1300\,\muHz$ after correcting for 
curvature using ridge centroids. The top panels show the diagrams calculated using the 
\most\ power spectrum, the middle panels using the 
radial velocity power spectrum, and the bottom panels using the product of both power 
spectra. Panels are separated
into the left-hand side ridge (panels a, c and e) and right-hand side ridge (panels b, d and e) seen 
in Figure \ref{fig06}. The dotted lines in the middle panels show the 
radial velocity power spectrum collapsed over the full oscillation range (corresponding to 
Figure 10 in \citet{bedding10})}
\label{fig07}
\end{center}
\end{figure}

Another possibility to search for ridge asymmetries is to increase the S/N by collapsing the 
\'{e}chelle diagram over several orders. To do so, we calculated the ridge centroids for the 
\most\ data in the same manner as done by \citet{bedding10} for the radial velocities. 
Using the ridge centroids calculated for the \most\ data, 
we straightened each order by removing the curvature seen in the \'{e}chelle diagram 
before collapsing the \'{e}chelle diagram in the frequency range where the centroids were reliably 
determined (700--1300\,\muHz). Figure \ref{fig07} shows a comparison with the analogous procedure for the velocity 
power spectrum \citep[see Figure 10 in][]{bedding10} as well as for the product of the \most\ and radial 
velocity power spectrum. Note that the ridge centroids have been calculated for each 
of the three power spectra individually since, as explained above, 
it is expected that the ridge positions are slightly different for each dataset. While the two 
ridges are clearly detected in the \most\ photometry (top panels), the S/N 
is too low to make any firm conclusion about the possible presence of separated ridges (i.e. 
$l=0$ and $l=2$), as is the case for the RV data (middle panels). We note, however, that the collapsed 
power spectrum of the combined data (bottom panels) might show some evidence for separated $l=0$ and $l=2$ components 
in the right-hand-side ridge. This would imply Scenario A, which is opposite to that preferred by 
\citet{bedding10} but in agreement with the results of Bayesian model comparisons using the RV data 
\citep{bedding10, handberg}. A more detailed analysis including the extraction of individual 
frequencies from the combined data and a comparison with pulsation models will be 
presented in a forthcoming paper (T. Kallinger et al., in preparation).

\subsection{Oscillation amplitudes}
\label{sec:amps}

\subsubsection{Amplitude Ratios}

The simultaneous observing campaigns allow us to measure the ratio of oscillation 
amplitudes in photometry and velocity and therefore test theoretical values and the scaling 
relations introduced by \citet{KB95}. The following influences have to be 
considered when measuring amplitudes of solar-like oscillations:

\begin{enumerate}
\item[(i)] Systematic variations due to the stellar cycle effects
\item[(ii)] Variations due to stochastic excitation and damping of the oscillation signal
\item[(iii)] Measurement errors due to background subtraction
\end{enumerate}

Since both datasets have been obtained within a timespan of less than 60\,days, 
effects arising from (i) can be safely ignored here. To measure amplitudes in a way that is 
largely insensitive to point (ii), we convolved the power spectrum with a Gaussian 
with FWHM = 4\Dnu\ and scaled the signal to the contribution of radial modes in each 
order \citep{kjeldsen08}. Note that for the factor $c$, which measures the effective number of modes 
per order, we have interpolated the values listed in 
Table 1 of \citet{kjeldsen08} to the central wavelength of the \most\ filter (525\,nm), 
yielding $c=3.14$. 

A crucial precondition to estimate the amplitude, particularly at low S/N, is to properly correct for 
the background contribution arising from stellar granulation and activity. In order to reliably 
estimate the uncertainty of the background parameters, we have used a combination of two 
published methods. An initial least-squares fit using the method of 
\citet{HSB09} was used as a starting point for a more careful fitting 
procedure using a Bayesian Markov-Chain Monte-Carlo (MCMC) algorithm, as described in \citet{gruberbauer09} 
and \citet{kallinger10}. The fitted model was adopted from \citet{karoff_phd} and has the form:

\begin{equation}
P(\nu) = P_{n} + \sum_{i=0}^k \frac{4\sigma_{i}^{2}\tau_{i}}{1+(2\pi\nu\tau_{i})^{2}+(2\pi\nu\tau_{i})^{4}} \: ,
\label{equ:harv}
\end{equation}

\noindent
where $P_{n}$ is the white noise component, $k$ is the number of power laws used and $\sigma$ and $\tau$ 
are the rms intensity and timescale of granulation, respectively. Note that in our application for 
Procyon, $k=2$.

\begin{figure}
\begin{center}
\resizebox{\hsize}{!}{\includegraphics{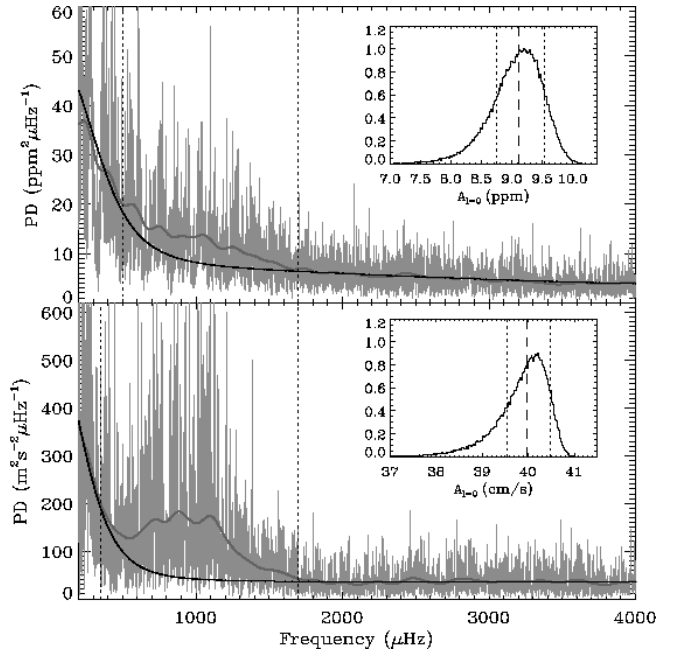}}
\caption{Power density spectra of Procyon smoothed with a 1\,\muHz\ boxcar (light grey) and 110\,\muHz\ boxcar (dark grey) 
observed in photometry (upper panel) and velocity 
(lower panel) with fitted background models (solid black lines). Vertical dotted lines mark the 
frequency intervals occupied by stellar oscillations which were excluded from the fit. The insets show 
the distributions of the mean amplitude per radial mode in the frequency interval 650-1150\,\muHz\ 
derived from the MCMC simulations of the background fit. Dashed lines mark the median and dotted lines the 
1-$\sigma$ confidence limits.}
\label{fig08}
\end{center}
\end{figure}

The determination of the resulting background and corresponding amplitude 
was done in two steps:

\begin{itemize}
\item[1)]  We excluded the region of the power spectrum which 
contains oscillation signal and ran the Bayesian MCMC algorithm only on the remaining power 
spectrum. The excluded region was determined by visual inspection of heavily 
smoothed power spectra (see Figure \ref{fig08}), and we verified through several trial MCMC runs 
that changing this region within reasonable limits does not significantly influence the results.
We used uniform priors for 
the granulation timescales and Jeffreys priors for the amplitudes of the background components. 
The most probable background model was then determined as the median 
of the marginalized posterior distributions for each background parameter.
\item[2)] Since the amplitude is not implicitly defined as a model parameter that is fitted to the data, 
we determined the pulsation amplitude as the mean level of the 
smoothed power spectrum in frequency range 650--1150\,\muHz\ after subtracting the background. 
This step was done for each MCMC iteration, and the final amplitude was then evaluated as the 
median and 1-$\sigma$ confidence limits 
of the resulting amplitude distribution.
\end{itemize}

Figure \ref{fig08} shows the background fits resulting from 10 independent MCMC chains, each with $10^{5}$ 
iterations, for the full \most\ and radial velocity datasets. The distributions 
for the mean amplitude over the range 
of 650--1150\,\muHz\ (after subtracting the background signal) yields 
$A_{l=0,\rm phot} = 9.1^{-0.4}_{-0.4}$\,ppm 
for the \most\ data and $A_{l=0,\rm RV}=40.0^{+0.5}_{-0.4}$\,cm$^{-1}$\,s for the velocity data. Hence, 
we arrive at a mean amplitude ratio for the full datasets of 
$A_{l=0,\rm phot}/A_{l=0,\rm RV} = 0.23 \pm 0.01$\,ppm\,cm$^{-1}$\,s.

As shown by \citet{kjeldsen08} for the Sun, short mode lifetimes can cause considerable variations 
even after
heavily smoothing the oscillation envelope.
To test these effects, we employed a similar approach as \citet{arentoft} and subdivided the 
\most\ data into five independent 
subsets of equal length ($\sim 7.5$\,d). Note that this subset length is considerably longer than  
the 10 individual 2\,d subsets used by \citet{arentoft} for the radial velocity data since the 
signal in photometry is lower, requiring longer sets 
to achieve a sufficiently high signal for an amplitude determination. The resulting smoothed 
curves, which were corrected for a background fit calculated using $5\times 10^5$ MCMC iterations for 
each subset, together with their mean value are shown in Figure \ref{fig09}(a). Using the central region 
of maximum power between 650-1150\,\muHz, we derived a mean amplitude of $A_{l=0,\rm phot} = 9.1\pm0.5$\,ppm,  
in very good agreement with the value derived using the full dataset above. 
In order to compare this photometric value to a velocity amplitude derived using the same 
method, we combine it with the estimate for the radial velocity amplitude given by \citet{arentoft} as 
$A_{l=0,\rm RV}=38.1\pm 1.3$\,cm$^{-1}$\,s. This yields an amplitude ratio, measured using subsets, of
$A_{l=0,\rm phot}/A_{l=0,\rm RV} = 0.24 \pm 0.02$\,ppm\,cm$^{-1}$\,s, 
again in good agreement with the value derived from the full datasets.

A few points need to be considered when evaluating the quoted values and uncertainties. Firstly,  
the fact that the uncertainty on $A_{l=0,\rm RV}$ derived from the full dataset is considerably 
lower than the value derived by \citet{arentoft} shows that for high S/N data the influence of 
stochastic excitation (see point (ii) above) dominates the amplitude uncertainty over the uncertainty 
arising from the background determination (see point (iii) above). For the \most\ data on the other 
hand, both uncertainties are roughly the same. To ensure a conservative approach, we  
therefore opted to use the amplitude ratio determined using subsets, 
$A_{l=0,\rm phot}/A_{l=0,\rm RV} = 0.24 \pm 0.02$\,ppm\,cm$^{-1}$\,s, as our final value of the 
mean amplitude ratio.

Secondly, the amplitude ratios do not include any uncertainties 
arising from the factors $c$ used to normalize the amplitude per radial mode. First observational 
constraints on photometric mode visibility ratios by \citet{deheuvels} suggest  $\sim2\sigma$ 
differences of up to 
25\% compared to theoretical responses used by \citet{kjeldsen08}. These differences could be caused by 
uncertainties in the limb-darkening laws used to calculate theoretical response functions, but also 
intrinsic differences in amplitudes of different degrees. We will assume that the latter cancel 
out in the photometry to velocity ratio, and therefore only concentrate on the mode visibilities.
We repeated the calculations of $c$ by numerically integrating the spatial response functions for 
a quadratic limb-darkening law for the Sun \citep{bedding96} assuming a conservative absolute uncertainty of 
0.05 (corresponding to a relative uncertainty of $\sim$10--20\%) for each limb-darkening coefficient 
\citep{howarth}. The resulting 
$c$ factors after 1000 integrations yield an uncertainty of 4\% in velocity and 6\% in photometry, 
which translates into uncertainties of 2\% and 3\% in the normalized amplitudes, and hence an 
uncertainty of 4\% in the amplitude ratio. This test shows that the uncertainty on $c$ can 
be substantial for estimating amplitude ratios, in particular for observations with high S/N and 
long mode lifetimes as found in red giant stars \citep{deridder}. In our case, however, the amplitude ratio is 
dominated by the $\sim$8\% uncertainty arising from the background fits and finite mode lifetimes.

\subsubsection{Comparison with Theoretical Results}

How do our estimates for the amplitude ratio compare with scaling relations? Rearranging 
equation (5) in \citet{KB95} yields

\begin{equation}
\frac{(\delta L/L)_{\lambda}/\rm ppm}{\nu_{\rm osc}/\rm cm\,s^{-1}} = \frac{20.1 \times 10^{-3}}{(\lambda/550 \rm nm) (\it{T}_{\rm eff}/\rm 5777\,\rm{K})^{\it r}} \: ,
\end{equation}

\noindent
with $r=1.5$ if the oscillations are adiabatic, and a best-fitting value of 
$r=2$ for observed amplitudes in classical pulsators \citep[see][]{KB95}.

\begin{figure}
\resizebox{8.5cm}{!}{\includegraphics{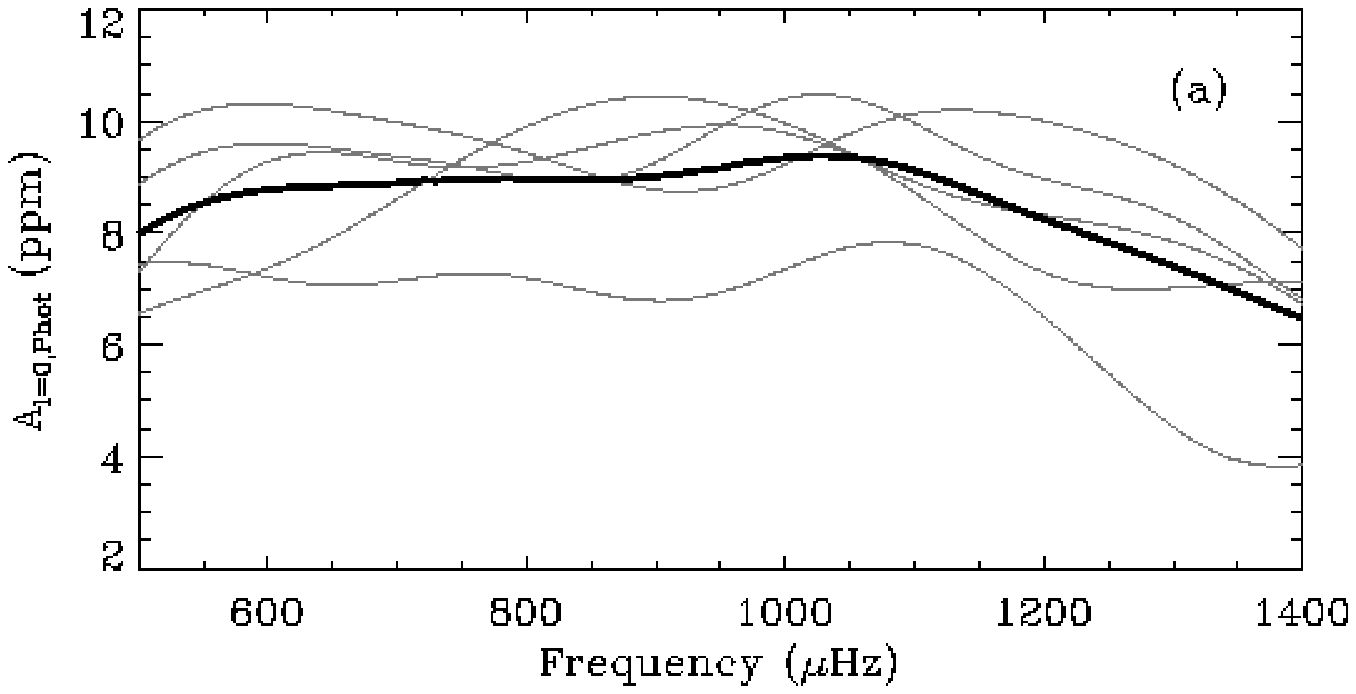}}
\resizebox{8.5cm}{!}{\includegraphics{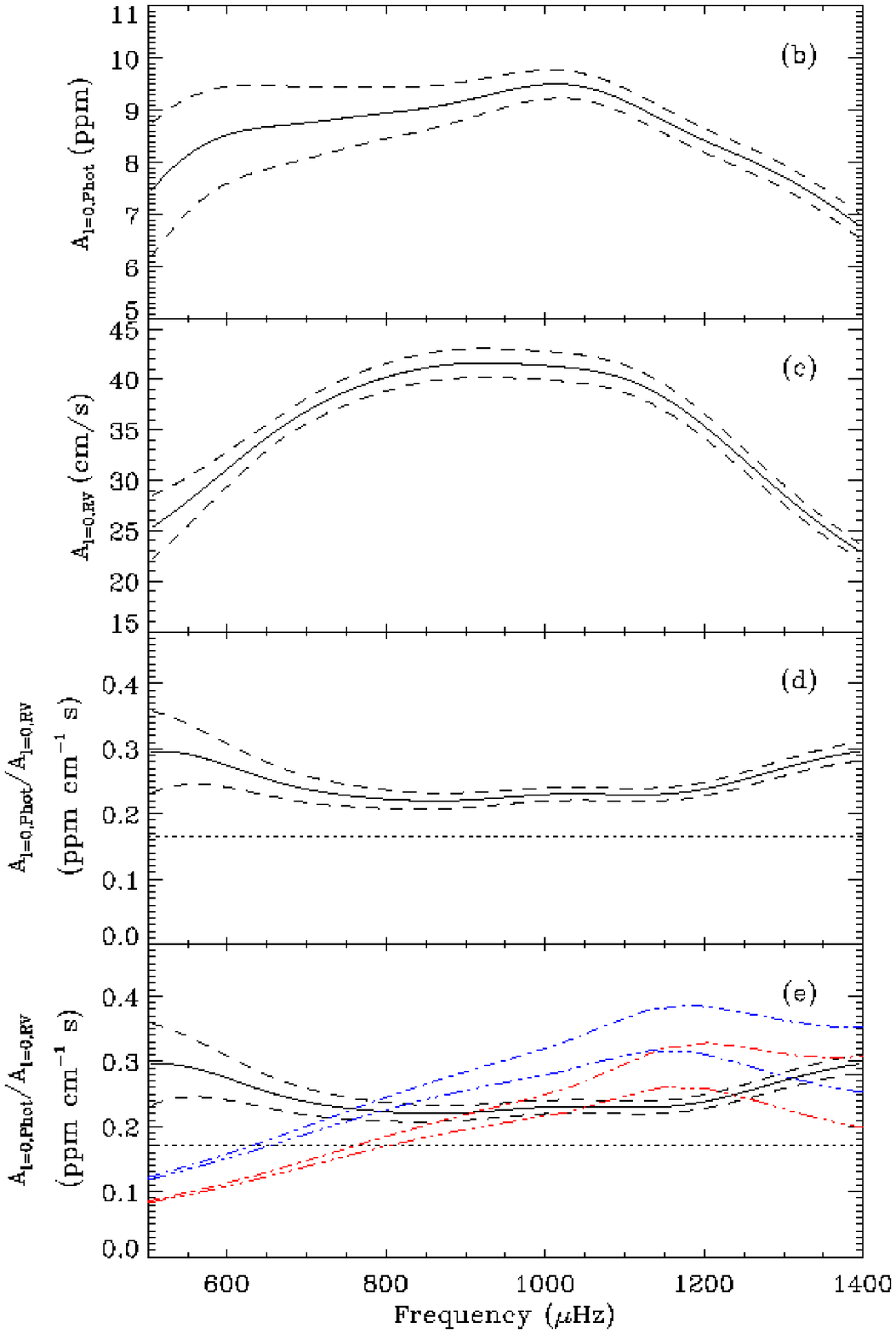}}
\caption{\textit{(a)} Smoothed amplitude curves of five independent 7.5\,d subsets of 
\most\ photometry (thin lines) together with their mean. \textit{(b)} Smoothed 
amplitude curve using the full \most\ dataset. Dashed lines mark the 1-$\sigma$ 
uncertainties. \textit{(c)} Same as panel (b) but for the full RV timeseries. 
\textit{(d)} Ratio of the curves shown in panels (b) and (c). 
The horizontal dotted line shows the expected amplitude ratio using the scaling relation of \citet{KB95}. 
\textit{(e)} Same as panel (d), but 
compared to theoretical amplitude ratios using an Eddington atmosphere (red lines) and a 
scaled VAL-C atmosphere (blue lines), for heights of 300\,km (dashed-dotted lines) and 600\,km 
(dashed-triple-dotted lines) above the photosphere \citep[see][for details]{houdek09}.}
\label{fig09}
\end{figure}

Using $r=2$, $T_{\rm eff}$ = 6500\,K and a wavelength $\lambda=525$\,nm, 
the expected amplitude ratio for radial modes is 0.17\,ppm\,cm$^{-1}$\,s. As shown by the dotted 
line in Figure \ref{fig09}(d), this is considerably lower than the observed ratio. Using $r=1.5$ the 
agreement is only slightly better, with an expected ratio of 0.18\,ppm\,cm$^{-1}$\,s.
Since the radial velocity amplitude is in agreement with the value from scaling relations 
\citep[see][]{arentoft}, this implies that the \most\ amplitude is higher than expected. 

Going one step further, the smoothed amplitude curves allow us to analyze the amplitude ratio as 
a function of frequency. Figure \ref{fig09}(d) shows the ratio of the amplitude curves 
derived from the full datasets for \most\ and the RV data, which are shown separately in Figure 
\ref{fig09}(b) and \ref{fig09}(c). Note that we have scaled the uncertainty of the RV 
amplitude curve to match the relative uncertainty derived using subsets, as described in the 
previous section. We also repeated the 
calculation using only HARPS data from the 
RV dataset, and restricted the \most\ data to the same timespan ($\sim$\,6.5\,d). The result was 
almost identical to the result based on the full dataset but with larger uncertainty. 

Figure \ref{fig09}(e) compares the amplitude ratio as a function of frequency with
theoretical predictions by \citet{houdek09} for a model of Procyon for different 
model atmospheres and different heights above the photosphere. Although the 
exact shape of the variation as a function of frequency is not well recovered, we note that 
the average amplitude ratio at maximum power ($\sim 1000\muHz$) is in better agreement with 
models than the estimate based on scaling relations (dotted line). A more detailed 
comparison will have to await the collection of higher S/N data from which amplitudes of 
individual mode frequencies can be reliably extracted.

It is interesting to note that, as can be seen from Figures \ref{fig06}, \ref{fig08}, \ref{fig09}(b) and 
\ref{fig09}(c), 
the maximum of the oscillation envelope in photometry seems to be shifted to slightly higher frequencies 
than in velocity. Defining \numax\ as the frequency corresponding to 
the maximum of the smoothed oscillation envelope, the MCMC analysis of the full datasets yields 
$\nu_{\rm max,phot}=1014^{+8}_{-11}\muHz$, compared to 
$\nu_{\rm max,RV}=923^{+9}_{-11}\muHz$. This shift translates into an increase of the amplitude ratio 
as a function of frequency, which tentatively can be identified in Figure \ref{fig09}(d). 

\section{Conclusions}

We have compared simultaneous space-based \most\ photometry 
and ground-based radial velocity data of the F5 star Procyon. Our main findings can be 
summarized as follows:

\begin{itemize}
\item The \most\ light curves of 2004, 2005 and 2007 show slow variations similar to those 
observed in velocity. A comparison of the variability level in photometry and velocity to the 
Sun confirmed that these variations are compatible with stellar activity on Procyon.

\item The peaks observed in the \most\ 2007 power spectrum match the oscillation frequencies 
detected in the radial velocity campaign, and the \'{e}chelle diagrams 
shows similar structure and curvature. We have attempted to 
confirm the mode ridge identification presented by \citet{bedding10} by 
collapsing power spectra corrected for curvature. 
While the \most\ data alone do not provide conclusive results, the collapsed \'{e}chelle diagram of the 
combined datasets shows some evidence that scenario A in \citet{bedding10} is the correct 
mode identification, contradicting the conclusions of that paper. Further work based on the extraction 
of individual frequencies from the combined data and a comparison with stellar models
will be necessary to confirm this result.

\item We measured the mean luminosity amplitude per radial mode in Procyon 
in the frequency range 650--1150\,\muHz\ to be 
$A_{l=0,\rm phot}= 9.1\pm0.5$\,ppm, 
in agreement with the value of $8.5\pm2$\,ppm published by \citet{bruntt}. Combining this with the 
mean velocity amplitude measured by \citet{arentoft} gives 
an amplitude ratio of $A_{l=0,\rm phot}/A_{l=0,\rm RV} = 0.24 \pm 0.02$\,ppm\,cm$^{-1}$\,s. This is 
considerably higher than the value of 0.17\,ppm\,cm$^{-1}$\,s expected from scaling from the Sun, but 
is in better
agreement with theoretical values predicted by \citet{houdek09}. We also analyzed the amplitude 
ratio as a function of frequency and found that the maximum of the 
oscillation envelope appears to be shifted to higher frequencies in photometry than in velocity.
\end{itemize}

The results presented here illustrate the potential of combining simultaneous luminosity and velocity 
measurements to study pulsations in stars. Future opportunities may arise from combining 
measurements from the ground-based radial-velocity network 
\textit{SONG} \citep{song} with space-based photometry by \most\ and \textit{BRITE} \citep{brite}, 
which will mark an important step in studying stellar structure and evolution in bright stars with 
well-known fundamental parameters.

\acknowledgments
The authors dedicate this publication to the memory of our dear friend and
colleague Dr.\ Piet Reegen, who sadly left us before his time.
We are thankful to our anonymous referee for very helpful comments which improved this manuscript. 
DH acknowledges support by the Astronomical Society 
of Australia (ASA). TRB and DS acknowledge support by the Australian Research Council.
GH acknowledges support by the Austrian FWF grant No. P2120521.
TK and WWW  are supported by the Austrian Fonds zur F\"orderung der wissenschaftlichen Forschung (FWF), 
project number P22691-N16. 
The Austrian participation in the \most\ project is funded by the Austrian Research Promotion Agency (FFG). 
MG, DBG and JMM acknowledge funding from the Natural Sciences \& Engineering Research Council 
(NSERC) Canada.

\bibliographystyle{apj}
\bibliography{references}

\end{document}